\documentclass[journal]{IEEEtran}
\usepackage{latexsym,,amssymb,amsmath,graphicx,epsf,cite,bbm,float}
\usepackage{ifpdf}
\usepackage{epstopdf,mathtools}

\usepackage{algorithm,algorithmic}
\usepackage{amsmath,amssymb,bm}
\usepackage{amsfonts,dsfont,color,bbm,subcaption}

\usepackage{mathtools}

\def\ninept{\def\baselinestretch{1}}
\ninept

\newcommand{\abs}[1]{|#1|}

\DeclareMathOperator*{\argmax}{arg\,max}
\DeclareMathOperator*{\argmin}{arg\,min}

\usepackage{hyperref,amsthm}

\newtheorem{theorem}{Theorem}

\newtheorem{lemma}[]{Lemma}

\newtheorem{proposition}[]{Proposition}

\newtheorem{definition}[]{Definition}

\begin{document}

\title{A 2-opt Algorithm for Locally Optimal\\ Set Partition Optimization} 
\author{\IEEEauthorblockN{Kaan Gokcesu}, \IEEEauthorblockN{Hakan Gokcesu} }
\maketitle

\begin{abstract}
	Our research deals with the optimization version of the set partition problem, where the objective is to minimize the absolute difference between the sums of the two disjoint partitions. Although this problem is known to be NP-hard and requires exponential time to solve, we propose a less demanding version of this problem where the goal is to find a locally optimal solution. In our approach, we consider the local optimality in respect to any movement of at most two elements. To accomplish this, we developed an algorithm that can generate a locally optimal solution in at most $O(N^2)$ time and $O(N)$ space. Our algorithm can handle arbitrary input precisions and does not require positive or integer inputs. Hence, it can be applied in various problem scenarios with ease.
\end{abstract}

\section{Introduction}\label{sec:intro}

\subsection{Set Partition Problem}
The set partition problem, as described in \cite{cook1971complexity,turing1937computable,levin1973universal,korf1998complete}, involves determining whether a given set $\mathcal{X}$ of positive integers can be partitioned into two complementary subsets $\mathcal{X}_1$ and $\mathcal{X}_2$, such that the sum of the elements in $\mathcal{X}_1$ is equal to the sum of the elements in $\mathcal{X}_2$. A related problem is the subset-sum problem \cite{kleinberg2006algorithm}, which aims to find a subset whose sum is equal to a target value $T$. This problem can be solved by adding a suitable dummy sample to the set and using a set partition solver.

In this paper, we investigate the optimization version of this problem, which involves partitioning the set $\mathcal{X}$ into two disjoint subsets $\mathcal{X}_1$ and $\mathcal{X}_2$, such that the absolute difference between the sum of elements in $\mathcal{X}_1$ and the sum of elements in $\mathcal{X}_2$ is minimized.

The traditional set partition problem is one of Karp's 21 NP-complete combinatorial problems \cite{karp1972reducibility} and one of Garey and Johnson's six NP-complete fundamental problems \cite{garey1979computers}. Similarly, the optimization version is NP-hard \cite{cieliebak2002equal,korf2009multi}. Although the traditional set partition problem is NP-complete, there still exist methods to solve it optimally or approximately for many instances. There are efficient pseudo-polynomial time algorithms that have a runtime polynomially dependent on the inputs.

The traditional set partition problem is often referred to as "the easiest hard problem" due to its limited structure in comparison to other NP-complete problems \cite{hayes2002computing,mertens2006number}. For example, the set of viable solutions can be exponentially large. It is a well-studied area of combinatorial optimization with significant developments \cite{graham1966bounds,coffman1978application,dell1995optimal,korf1998complete,korf2009multi,moffitt2013search,schreiber2013improved,schreiber2014cached,schreiber2018optimal}. The study of this problem may provide insight into potential improvements for harder NP-complete problems \cite{schreiber2018optimal}.

\subsection{Example Applications}
The set partitioning problem has a wide range of applications in various fields such as learning, optimization, and decision theory. These applications include but are not limited to scheduling, allocation, classification and regression \cite{gokcesu2021nonparametric,cano2007evolutionary,gokcesu2020recursive,golbraikh2000predictive,gokcesu2020generalized,fan2005working,huberbook,cesabook,poor_book,kellerer2004knapsack,gokcesu2022nonconvex,mathews1896partition,dantzig2007number,coffman1984approximation,brams1996fair,biswas2018fair,gokcesu2021optimal,walsh2009really,merkle1978hiding,shamir1982polynomial,rivest1983cryptographic,gokcesu2021optimally,sarkar1987partitioning,dell2008heuristic,graham1979optimization,umang2013exact,lalla2016set,gokcesu2022low}.

One closely related problem is the knapsack problem \cite{kellerer2004knapsack}, where the objective is to determine the number of each item in a collection so that the total weight is constrained by an upper bound, and the total value is maximized. This problem often arises in resource allocation settings, where we need to select non-divisible tasks or goods under a time or budget constraint. The knapsack problem has been a century-old problem \cite{mathews1896partition}, and its name dates back to the early works of Dantzig \cite{dantzig2007number}. It has numerous applications in real-world decision-making processes across different fields, such as finding the least wasteful way to cut raw materials, selecting investments and portfolios, selecting assets for securitization, and generating keys for the Merkle–Hellman and other knapsack cryptosystems \cite{kellerer2004knapsack}.

The bin packing optimization problem \cite{coffman1984approximation} is another closely related problem. The objective of this problem is to pack items with varying sizes into a minimum number of bins with  a predetermined fixed capacity. It has numerous applications, from filling containers to loading trucks and creating file backups.

Another related problem is the fair division problem \cite{brams1996fair}, where the goal is to fairly divide a set of resources among a certain number of recipients. This problem arises in many real-world scenarios, such as division of inheritance, dissolution of partnerships, settlements in divorces, frequency allocation in band usage, traffic management of airports or harbors, satellite allocations, and social choice theory \cite{brams1996fair,biswas2018fair}.

The partition problem is synonymous with parallel program scheduling in some works \cite{dell2008heuristic}, where the objective is to assign various tasks to parallel working machines while minimizing the total amount of time it takes for their completion \cite{garey1979computers,sarkar1987partitioning,dell2008heuristic,graham1979optimization}. A similar application is in berth allocation \cite{umang2013exact}, which aims to optimally allocate berth space for incoming vessels in container terminals while minimizing the total time it takes to serve all the vessels \cite{umang2013exact,lalla2016set}.

Fair team selection is another well-known problem related to set partitions. The objective is to create fair teams regarding the total skill level of the players, i.e., minimizing the difference between the cumulative skill of players in different teams \cite{hayes2002computing}. 

Another example is the manipulation of veto elections, where each elector has different veto weights, and the candidate with the smallest total veto wins the election. Using set partitioning, a coalition can maximize their candidate's chances \cite{walsh2009really}.

\subsection{Existing Algorithms}
Solving the set partition problem for an input set $\mathcal{X}$ of size $N$ is known to be an NP-hard problem, with a straightforward brute-force approach having a time complexity of $O(2^N)$ \cite{garey1979computers}. However, more efficient algorithms have been developed to address this issue. Horowitz and Sahni proposed an algorithm in 1974 that finds an optimal solution in $O(2^{N/2})$ time but requires $O(2^{N/2})$ memory \cite{horowitz1974computing}. This exponential memory usage was later addressed by Schroeppel and Shamir in 1981, who decreased the space complexity to $O(2^{N/4})$ while keeping the same time complexity \cite{schroeppel1981t}.

In addition to exponential-time approaches, dynamic programming algorithms \cite{garey1979computers,martello1990knapsack,korf2013optimally} have also been developed to solve the set partition problem. These algorithms find an optimal solution in pseudo-polynomial time and space, which are polynomially dependent on the sum of the inputs $S$. However, their performance is highly dependent on the precision of the inputs, which must be positive and integer.

Sub-optimal algorithms such as the greedy algorithm and the Karmarkar-Karp set differencing algorithm have also been developed, which run in $O(N\log N)$ time and $O(N)$ space \cite{graham1966bounds,karmarkar1982differencing,kellerer2004knapsack,korf2011hybrid}. These algorithms are useful when an optimal solution is not necessary, and a good approximation is sufficient. Locally optimal efficient algorithms have also been proposed for distinct scenarios \cite{gokcesu2021efficient,gokcesu2021quadratic,gokcesu2022linearithmic}.

Additionally, the complete anytime algorithm by Korf in 1998 can transform sub-optimal algorithms into optimal ones and create anytime algorithms with linear memory usage \cite{korf1998complete}. However, the worst-case time complexity of this approach is $O(2^N)$.

\subsection{Contributions}
The set partition problem is a well-known problem in computer science, and in this paper, we aim to tackle the issue of finding efficient algorithms to solve it. 

Despite the existence of optimal algorithms, their exponential time and space complexities make them impractical. 

Dynamic programming approaches offer optimal solutions with a pseudo-polynomial complexity, but they are limited in their use for non-integer or high precision inputs.

There exist linearithmic time complexity algorithms that produce sub-optimal solutions, and the existing locally optimal algorithms have weak local optimality.

To address these challenges, we propose a novel algorithm that finds a locally 2-optimal solution to the set partition problem in $O(N^2)$ time and $O(N)$ space. Our approach postulates that finding a locally optimal solution is considerably easier than finding a globally optimal one.

\subsection{Organization}
The paper is organized as follows: 
\begin{itemize}
	\item In \autoref{sec:prob}, we present the mathematical formulation of the problem. 
	\item In \autoref{sec:method}, we describe our proposed algorithm and provide some important results. \item In \autoref{sec:analyses}, and provide our algorithm's time and space complexities. 
	\item In \autoref{sec:disc}, we conclude with some remarks.
\end{itemize}

\section{Locally 2-Optimal Set Partition Problem}\label{sec:prob}
We aim to formally define the 2-locally optimal set partition problem. Given a set of numbers $$\mathcal{X}=\{x_1, x_2, \ldots, x_N\}=\{x_n\}_{n=1}^N,$$ where $N$ is an integer, the problem involves partitioning $\mathcal{X}$ into two subsets $\mathcal{X}_1$ and $\mathcal{X}_2$, such that they are disjoint and their union is equal to $\mathcal{X}$. The cardinalities of $\mathcal{X}_1$ and $\mathcal{X}_2$ add up to $N$, i.e.,
\begin{align}
	\mathcal{X}_1\cap\mathcal{X}_2&=\emptyset,\\
	\mathcal{X}_1\cup\mathcal{X}_2&=\mathcal{X},\\
	\abs{\mathcal{X}_1}+\abs{\mathcal{X}_2}&={N},
\end{align}	

We are interested in creating the sets $\mathcal{X}_1$ and $\mathcal{X}_2$ such that their individual set sums $S_1$ and $S_2$, respectively, are as close as possible to each other. The set sums as follows:
\begin{align}
	S_1 &= \sum_{x\in\mathcal{X}_1}x \\
	S_2 &= \sum_{x\in\mathcal{X}_2}x \\
	S &= \sum_{x\in\mathcal{X}}x,
\end{align}
where $S_1$ and $S_2$ are complementary, such that $S_1 = S - S_2$ and $S_2 = S - S_1$. We aim to minimize the absolute difference between the sums $S_1$ and $S_2$, which can be formulated as:
\begin{align}
	\min_{\mathcal{X}_1,\mathcal{X}_2} \left(\abs{S_1-S_2}\right).
\end{align}
Alternatively, we can maximize the minimum sum or minimize the maximum sum. However, all of these formulations are equivalent to each other when dealing with two sets and their corresponding sums.

This problem is known to be NP-hard and cannot be solved with an efficient method. To tackle this issue, we focus on a weaker version of the set partition problem, which aims to obtain a locally optimal solution instead of a global one.

\begin{definition}\label{def:opt}
A partition $\mathcal{X}_1,\mathcal{X}_2$ is a locally 2-optimal partition if there does not exist a pair of elements $x_1, x_2 \in \mathcal{X}\cup\{0\}$, where moving them to the other set would decrease the absolute difference between the set sums $S_1$ and $S_2$. Formally, we have:
\begin{align}
	\abs{\tilde{S}_1-\tilde{S}_2}\geq\abs{S_1-S_2},
\end{align}
where $\tilde{S}_1$ and $\tilde{S}_2$ represent the new set sums obtained by moving $x_1$ and $x_2$ to the other set, i.e.,
\begin{align*}
	\tilde{S}_1=&S_1-\mathbbm{1}_{x_1\in\mathcal{X}_1}x_1-\mathbbm{1}_{x_2\in\mathcal{X}_1}x_2+\mathbbm{1}_{x_1\in\mathcal{X}_2}x_1+\mathbbm{1}_{x_2\in\mathcal{X}_2}x_2,
	\\\tilde{S}_2=&S_2-\mathbbm{1}_{x_1\in\mathcal{X}_2}x_1-\mathbbm{1}_{x_2\in\mathcal{X}_2}x_2+\mathbbm{1}_{x_1\in\mathcal{X}_1}x_1+\mathbbm{1}_{x_2\in\mathcal{X}_1}x_2,
\end{align*}
for all $x_1,x_2\in\mathcal{X}\cup\{0\}$.
\end{definition}

In the next section, we propose an algorithm that can reach such a locally optimal solution in quadratic time.

\section{Finding a Locally 2-Optimal Solution to the Set Partition Problem}\label{sec:method}
In this section, we provide the algorithm and make some important remarks, which will be used in our analysis in the next section.
Given the input set ${\mathcal{X}}$, the algorithm works as:
\begin{enumerate}
	\item Append the set with $N$ zeros to create an extended set of size $2N$, i.e., 
	\begin{align*}
		\mathcal{X}_E=\mathcal{X}\cup\{x_{n}\}_{n=N+1}^{2N},
	\end{align*}
	where $x_n=0$ for $n\in\{N+1,\ldots,2N\}$.
	\item Create an indicator set $\mathcal{I}=\{i_n\}_{n=1}^{2N}$ from $\mathcal{X}_E$, where
	\begin{align}
		i_n=\begin{cases}
			1,& {x}_n>0\\
			0,& x_n=0\\
			-1.& {x}_n< 0
		\end{cases}
	\end{align}
	Hence, this set contains the sign of every element. Obviously $i_n=0$ for $n\in\{N+1,\ldots,2N\}$.
	\item Create an absolute set ${\mathcal{X}}_{A}=\{{x}^A_n\}_{n=1}^{2N}$, where ${x}^A_n=\abs{x_n}$ for all $n$. Hence, $$x_n=i_n{x}^A_n.$$
	\item Start with an arbitrary equal sized partition $\mathcal{X}_1$ and $\mathcal{X}_2$ of ${\mathcal{X}}_A$ with the respective initial sums $S_1$ and $S_2$, i.e., $\abs{\mathcal{X}_1}=\abs{\mathcal{X}_2}=N$\label{step:start}.
	\item If $S_1=S_2$, skip to Step \ref{step:final}\\
			else, set 
			\begin{align*}
				a=&\argmax_{i\in\{1,2\}}S_i,
				\\
				b=&\argmin_{i\in\{1,2\}}S_i.
			\end{align*}
	Hence, the set $\mathcal{X}_a$ has a larger sum then $\mathcal{X}_b.$ \label{step:n}		
	\item Set $n=1$.
	\item Let $x_a^n$ be the $n^{th}$ largest element of $\mathcal{X}_a$.
	\item Check if swapping $x_a^n$ with any element $x_b<x_a^n$ from $\mathcal{X}_b$ strictly decreases the sum difference $\abs{S_1-S_2}$. \label{step:xn}
	\item If yes, continue; \\
	else if $n=N$, skip to Step \ref{step:final};
	\\ or else, set $n\leftarrow n+1$ and return to Step \ref{step:xn}.\label{step:incn}
	\item Swap $x_a^n\in\mathcal{X}_a$ with the smaller $x_b\in\mathcal{X}_b$ that most decreases the absolute set sum difference, i.e.,
	\begin{align*}
		x_b=\argmin_{x<x_a^n, x\in\mathcal{X}_b}|S_a-S_b-2{x}^a_n+2x|.
	\end{align*}
	Do the necessary adjustments
	\begin{align*}
		\mathcal{X}_a&\leftarrow (\mathcal{X}_a\setminus\{{x}^a_n\})\cup\{x_b\},
		\\ \mathcal{X}_b&\leftarrow (\mathcal{X}_b\setminus\{x_b\})\cup\{x^a_{n}\},
		\\ S_a&\leftarrow S_a-{x}^a_n+x_b,
		\\ 
		S_b&\leftarrow S_b-x_b+x_a^n,
	\end{align*}
	\item If the sign of $S_1-S_2$ remains unchanged,
	\subitem if $n=N$, continue;
	\subitem or else, set $n\leftarrow n+1$ and return to Step \ref{step:xn};
	\\or else, return to Step \ref{step:n}.\label{step:it}
	\item Move each ${x}^A_n$ to the opposing set if $i_n=-1$.\label{step:final}
	\item Set each ${x}^A_n\leftarrow i_n{x}^A_n$. Remove the redundant zeros added in the beginning such that $\abs{\mathcal{X}_1}+\abs{\mathcal{X}_2}=N$.
\end{enumerate}

 
 \begin{proposition}
 	The algorithm checks all possible swaps of $x_a^n\in\mathcal{X}_a$ with any $x\in\mathcal{X}_b$ that can strictly decrease the sum difference $\abs{S_a-S_b}$.
 	\begin{proof}
 		As in the algorithm, let $S_a-S_b>0$. For the absolute sum difference $\abs{S_1-S_2}=\abs{S_a-S_b}$ to decrease, the swapped elements $x_a^n$ and $x_b$ should have the inequality $x_a^n>x_b$. Otherwise, $S_a-S_b$ will get larger and so is $\abs{S_a-S_b}=\abs{S_1-S_2}$.
 	\end{proof}
 \end{proposition}

 \begin{lemma}\label{thm:SoldSnew}
 	As in the algorithm, let $S_a^{(1)}-S_b^{(1)}=\triangle^{(1)}>0$, at some point in the algorithm. If swapping the elements $x_{a}\in\mathcal{X}_a$ and $x_{b}\in\mathcal{X}_b$ ($x_a>x_b$) did not decrease the absolute sum difference $\abs{S_a-S_b}$; neither does swapping them when $S_a^{(2)}-S_b^{(2)}=\triangle^{(2)}>0$, where $\triangle^{(2)}\leq\triangle^{(1)}$.
 	\begin{proof}
 		Since the swap did not decrease the absolute sum difference $\abs{S_a-S_b}$, we have $x_a-x_b\geq \triangle^{(1)}$, which also implies $x_a-x_b\geq \triangle^{(2)}$ and concludes the proof.
 	\end{proof}
 \end{lemma}
 
 \begin{lemma}\label{thm:xn2}
 	As in the algorithm, let $S_a-S_b>0$. Let $x_a^n\in\mathcal{X}_a$, $x_{b}\in\mathcal{X}_b$ (such that $x_a^n>x_b$) be a pair of elements whose swap maximally decreases the absolute sum difference $\abs{S_a-S_b}$ over all possible swaps including $x_{a}^n$. After the swap, there are no elements $x'_b\in\mathcal{X}_2$ such that $x'_b\leq x_b$, whose swap with $x_{b}$ strictly decreases the absolute difference of set sums.
 	\begin{proof}
 		The proof comes from the fact that the pair $x_a^n$ and $x_b$ is a minimizer swap over the swaps including $x_a^n$. If there indeed were a feasible $x'_b$, the minimizer swap would simply be between $x_a^n$ and $x'_b$.
 	\end{proof}
 \end{lemma}

 \begin{lemma}\label{thm:X2}
	In the algorithm, before a swap happens between $x_a^n\in\mathcal{X}_a$ and $x_b\in\mathcal{X}_b$, there exists no $x'_a\in\mathcal{X}_a$ such that $x_b<x'_a<x_a^n$.
	\begin{proof}
		We prove by contradiction. Let us assume there exists $x_b<x'_a<x_a^n$ in the set $\mathcal{X}_a$. The algorithm checks $x_a^n\in\mathcal{X}_a$ only if this $x'_a\in\mathcal{X}_a$ has no viable operations. 
		Since $x'_a>x_b$ and $x'_a$ has no viable operations, we have $x'_a-x_b>S_a-S_b>0$. However, since $x'_a<x_a^n$, we also have $x_a^n-x_b>S_a-S_b$. Thus, the swap of $x_a^n$ and $x_b$ is also not a viable operation since after their swap the absolute sum difference would increase. This contradicts with the initial situation that stated the swap of $x_a^n$ and $x_b$. Hence, there cannot exist a $x'_a\in\mathcal{X}_a$ such that $x_b<x'_a<x_a^n$, which concludes the proof.
	\end{proof}
\end{lemma}

\begin{lemma}\label{thm:decAbs}
	In the algorithm, whenever a sign change happens on $S_1-S_2$ after a swap between $x_{a}^n\in\mathcal{X}_a$ and $x_{b}\in\mathcal{X}_b$; we have $\abs{S_a-S_b-2x_{a}^n+2x_{b}}\leq \abs{\min_{x>x_b, x\in\mathcal{X}_A}x-x_b}$.
	\begin{proof}
		From \autoref{thm:X2}, we have that whenever there is a swap between $x_{a}^n\in\mathcal{X}_a$ and $x_{b}\in\mathcal{X}_b$, all the elements between them are in $\mathcal{X}_b$, i.e., if $x_b<x<x_a^n$ for any $x\in\mathcal{X}_A$, we have $x\in\mathcal{X}_b$. Thus, this swap is equivalent to swapping all $x_b< x< x_a^n$ consecutively in decreasing order. For example, let there exist two elements such that $x_b<x_1\leq x_2< x_a^n $. Since $x_a^n\in\mathcal{X}_a$ and $x_b,x_1,x_2\in\mathcal{X}_b$; swapping $x_a^n$ and $x_b$ is equivalent to swapping $(x_a^n,x_2)$, $(x_2,x_1)$, $(x_1,x_b)$ in order. Thus, any swap in the algorithm can be expressed as a series of swaps between consecutive elements.	Since a series of consecutive swaps results in the sign change. The last swap in the series results in the sign change, which concludes the proof.
	\end{proof}
\end{lemma}

\section{Analyses of the Algorithm and Its Output}\label{sec:analyses}
In this section, we analyze our algorithm by iteratively showing that:
\begin{enumerate}
	\item our algorithm terminates and reaches a solution partition $\mathcal{X}_1,\mathcal{X}_2$ with at most $O(N^2)$ time and $O(N)$ space complexity.
	\item when our algorithm terminates, the produced partitioning $\mathcal{X}_1,\mathcal{X}_2$ is a locally 2-optimal solution as in \autoref{def:opt}.
\end{enumerate}


We start our analyses of our algorithm by first showing its computational complexity.
 \begin{theorem}
 	The algorithm reaches a solution partition $\mathcal{X}_1,\mathcal{X}_2$ in at most $O(N^2)$ time and $O(N)$ space complexity.
 	\begin{proof}
 		As in the algorithm, we have $S_a-S_b>0$ in any of our traverse from $n=1$ to $n=N$ for some suitable $a,b$ (note that whenever the sign changes we begin a new traverse). Suppose we are at some time iteration $n_1$. Let $$\max_{x\in\mathcal{X}_a:x<x_a^{n_1}}x=x_a^{n_2}.$$ This implies that any $x\in\mathcal{X}_a$ such that $x\leq x_a^{n_2}$ has no viable swaps with any element $x'\in\mathcal{X}_b$. From \autoref{thm:X2}, we know that $x_a^{n_1}$ can only be swapped with some $x\in\mathcal{X}_b$ such that $$x_a^{n_2}\leq x< x_a^{n_1},$$ and there exists no $x'\in\mathcal{X}_a$ such that $x_a^{n_2}< x'< x_a^{n_1}$. Suppose there exists $n'$ number of such elements in $\mathcal{X}_b$, i.e.,
 		$$\abs{\mathcal{X}_b\cap[x_a^{n_2},x_a^{n_1})}=n'.$$
 		To find the best swap, we start by checking the swap with the smallest one, i.e., 
 		$$\min_{x\in\mathcal{X}_b:x_a^{n_2}\leq x<x_a^{n_1}}x,$$
 		and do a simple linear search, which takes at most $O(n')$ time. In a single traverse, in the worst-case, we iteratively do this for all the elements in $\mathcal{X}_a$, hence, it takes at most $O(N)$ time. Thus, each traverse over $n$ (i.e., from $n=1$ to $n=N$) takes $O(N)$ time.
 		We point out that a traverse ends after a swap which causes a sign change in $S_1-S_2$. From \autoref{thm:decAbs}, we have that whenever the sign changes, the resulting absolute difference between the set sums will be less than the difference between some consecutive elements in $\mathcal{X}_A$. Since with each swap, the absolute difference decreases, this consecutive swap can never happen again after it has happened already once. Because there exist $N-1$ possible consecutive element pairs, the number of traverses can never exceed $N$, i.e., we have at most $N$ traverses. Since we have $N$ traverses and each traverse takes $O(N)$ time, we have at most $O(N^2)$ time complexity. Since our algorithm only keeps track of which elements are in which subset, it uses $O(N)$ space, which concludes the proof. 
 	\end{proof}
 \end{theorem}
 
 We have shown that our algorithm terminates and produces a partition in at most $O(N^2)$ time and space complexity. Next, we conclude our analyses by investigating the local optimality of the produced partition.
 \begin{theorem}
 	The algorithm's solution partition $\mathcal{X}_1,\mathcal{X}_2$ is locally 2-optimal as in \autoref{def:opt}.
 	\begin{proof}
 		The algorithm terminates when Step \ref{step:final} is reached. If we reached it from Step \ref{step:n}, there exist no pairs $x_1\in\mathcal{X}_1$ and $x_2\in\mathcal{X}_2$ whose swap decreases the absolute difference of the set sums since $\abs{S_1-S_2}=0$. If we reach it from Step \ref{step:incn} or Step \ref{step:it}, we make a final traverse from $n=1$ to $n=N$ all the while never returning to Step \ref{step:n} from Step \ref{step:it}. In this traverse, $n$ increases by $1$, if swapping $x_a^n$ with any $x<x_a^n$ is unfruitful or we make a swap and the sign of $S_1-S_2$ remains unchanged. If there are no swaps in the traverse, that means there are no pairs $x_{a}\in\mathcal{X}_a$ and $x_{b}\in\mathcal{X}_b$ ($x_a>x_b$) whose swap decreases the absolute difference of set sums.
 		Suppose there is a swap between $x_{a}\in\mathcal{X}_a$ and $x_{b}\in\mathcal{X}_b$ ($x_a>x_b$). After the swap, from \autoref{thm:SoldSnew}, we have that $x\in\mathcal{X}_a$ for $x<x_a$ and $x\neq x_b$ still has no viable swaps. Moreover, from \autoref{thm:xn2}, we have that $x_b$ has no viable swap. Thus, $x\in\mathcal{X}_1$ for $x<x_a$ has no viable swap. Henceforth, in the final traverse, the swaps eliminate any remaining viable swaps while not affecting the smaller elements. At $n=N$, we reach a situation where there exists no swaps that decreases the absolute difference of the set sums, i.e., no pairs $x_1\in\mathcal{X}_1$ and $x_2\in\mathcal{X}_2$ whose swap decreases the absolute difference of set sums $\abs{S_1-S_2}$. From the $N$ original elements and $N$ redundant zeros, there exists in total $2N$ elements which are equally divided between $\mathcal{X}_1$ and $\mathcal{X}_2$ at the end. Either all redundant zeros are at one of the sets or both sets include at least one redundant zero; which means for every nonzero element in $\mathcal{X}_1$ there exists a zero in $\mathcal{X}_2$ and vice-versa. Since all nonzero elements are positive just before Step \ref{step:final}, and there exists no viable swaps, moving two nonzero elements together from one set to the another also not viable. Hence, we reach a situation where swapping two nonzero elements and moving one or two nonzero elements  does not decrease $\abs{S_1-S_2}$. Since a swap is equivalent to moving two element separately, there does not exist any one or two moves that decrease $\abs{S_1-S_2}$. We point out that moving an element from its set to the opposing set is equivalent to the converse if the element were to be multiplied by $-1$. Hence, at Step \ref{step:final} we move such elements with indicator $-1$ to the opposing set and restore their original values. After removing the redundant zeros, we reach a partition $\mathcal{X}_1,\mathcal{X}_2$ such that there does not exists at most two elements whose set change decreases the absolute set sum difference, which is equivalent to our local optimality criterion in \autoref{def:opt}.
 	\end{proof}
 \end{theorem}

\section{Conclusion}\label{sec:disc}

In conclusion, we have studied the optimization version of the set partition problem where the absolute difference between the partition sums are minimized. Although this problem is NP-hard, we have formulated a weaker version, where the goal is to find a locally 2-optimal solution. The local optimality considered in our work is under at most two movement of the elements. To this end, we designed an algorithm which can produce such a locally optimal solution in $O(N^2)$ time and $O(N)$ space. Our approach does not require positive, integer inputs; and works equally well under arbitrary input precisions, which makes it widely applicable.

\bibliographystyle{IEEEtran}
\bibliography{double_bib}

\begin{thebibliography}{10}
\providecommand{\url}[1]{#1}
\csname url@samestyle\endcsname
\providecommand{\newblock}{\relax}
\providecommand{\bibinfo}[2]{#2}
\providecommand{\BIBentrySTDinterwordspacing}{\spaceskip=0pt\relax}
\providecommand{\BIBentryALTinterwordstretchfactor}{4}
\providecommand{\BIBentryALTinterwordspacing}{\spaceskip=\fontdimen2\font plus
\BIBentryALTinterwordstretchfactor\fontdimen3\font minus
  \fontdimen4\font\relax}
\providecommand{\BIBforeignlanguage}[2]{{%
\expandafter\ifx\csname l@#1\endcsname\relax
\typeout{** WARNING: IEEEtran.bst: No hyphenation pattern has been}%
\typeout{** loaded for the language `#1'. Using the pattern for}%
\typeout{** the default language instead.}%
\else
\language=\csname l@#1\endcsname
\fi
#2}}
\providecommand{\BIBdecl}{\relax}
\BIBdecl

\bibitem{cook1971complexity}
S.~A. Cook, ``The complexity of theorem-proving procedures,'' in
  \emph{Proceedings of the third annual ACM symposium on Theory of computing},
  1971, pp. 151--158.

\bibitem{turing1937computable}
A.~M. Turing, ``On computable numbers, with an application to the
  entscheidungsproblem,'' \emph{Proceedings of the London mathematical
  society}, vol.~2, no.~1, pp. 230--265, 1937.

\bibitem{levin1973universal}
L.~A. Levin, ``Universal sequential search problems,'' \emph{Problemy peredachi
  informatsii}, vol.~9, no.~3, pp. 115--116, 1973.

\bibitem{korf1998complete}
R.~E. Korf, ``A complete anytime algorithm for number partitioning,''
  \emph{Artificial Intelligence}, vol. 106, no.~2, pp. 181--203, 1998.

\bibitem{kleinberg2006algorithm}
J.~Kleinberg and E.~Tardos, \emph{Algorithm design}.\hskip 1em plus 0.5em minus
  0.4em\relax Pearson Education India, 2006.

\bibitem{karp1972reducibility}
R.~M. Karp, ``Reducibility among combinatorial problems,'' in \emph{Complexity
  of computer computations}.\hskip 1em plus 0.5em minus 0.4em\relax Springer,
  1972, pp. 85--103.

\bibitem{garey1979computers}
M.~R. Garey and D.~S. Johnson, \emph{Computers and intractability}.\hskip 1em
  plus 0.5em minus 0.4em\relax freeman San Francisco, 1979, vol. 174.

\bibitem{cieliebak2002equal}
M.~Cieliebak, S.~J. Eidenbenz, A.~Pagourtzis, and K.~Schlude, ``Equal sum
  subsets: complexity of variations,'' \emph{Technical Report/ETH Zurich,
  Department of Computer Science}, vol. 370, 2002.

\bibitem{korf2009multi}
R.~E. Korf, ``Multi-way number partitioning,'' in \emph{Twenty-First
  International Joint Conference on Artificial Intelligence}, 2009.

\bibitem{hayes2002computing}
B.~Hayes, ``Computing science: The easiest hard problem,'' \emph{American
  Scientist}, vol.~90, no.~2, pp. 113--117, 2002.

\bibitem{mertens2006number}
S.~Mertens, ``Number partitioning,'' \emph{Computational Complexity and
  Statistical Physics}, p. 125, 2006.

\bibitem{graham1966bounds}
R.~L. Graham, ``Bounds for certain multiprocessing anomalies,'' \emph{Bell
  system technical journal}, vol.~45, no.~9, pp. 1563--1581, 1966.

\bibitem{coffman1978application}
E.~G. Coffman, Jr, M.~R. Garey, and D.~S. Johnson, ``An application of
  bin-packing to multiprocessor scheduling,'' \emph{SIAM Journal on Computing},
  vol.~7, no.~1, pp. 1--17, 1978.

\bibitem{dell1995optimal}
M.~Dell’Amico and S.~Martello, ``Optimal scheduling of tasks on identical
  parallel processors,'' \emph{ORSA Journal on Computing}, vol.~7, no.~2, pp.
  191--200, 1995.

\bibitem{moffitt2013search}
M.~D. Moffitt, ``Search strategies for optimal multi-way number partitioning,''
  in \emph{Twenty-Third International Joint Conference on Artificial
  Intelligence}, 2013.

\bibitem{schreiber2013improved}
E.~L. Schreiber and R.~E. Korf, ``Improved bin completion for optimal bin
  packing and number partitioning,'' in \emph{Twenty-Third International Joint
  Conference on Artificial Intelligence}, 2013.

\bibitem{schreiber2014cached}
E.~Schreiber and R.~Korf, ``Cached iterative weakening for optimal multi-way
  number partitioning,'' in \emph{Proceedings of the AAAI Conference on
  Artificial Intelligence}, vol.~28, no.~1, 2014.

\bibitem{schreiber2018optimal}
E.~L. Schreiber, R.~E. Korf, and M.~D. Moffitt, ``Optimal multi-way number
  partitioning,'' \emph{Journal of the ACM (JACM)}, vol.~65, no.~4, pp. 1--61,
  2018.

\bibitem{gokcesu2021nonparametric}
K.~Gokcesu and H.~Gokcesu, ``Nonparametric extrema analysis in time series for
  envelope extraction, peak detection and clustering,'' \emph{arXiv preprint
  arXiv:2109.02082}, 2021.

\bibitem{cano2007evolutionary}
J.~R. Cano, F.~Herrera, and M.~Lozano, ``Evolutionary stratified training set
  selection for extracting classification rules with trade off
  precision-interpretability,'' \emph{Data \& Knowledge Engineering}, vol.~60,
  no.~1, pp. 90--108, 2007.

\bibitem{gokcesu2020recursive}
K.~Gokcesu and H.~Gokcesu, ``Recursive experts: An efficient optimal mixture of
  learning systems in dynamic environments,'' \emph{arXiv preprint
  arXiv:2009.09249}, 2020.

\bibitem{golbraikh2000predictive}
A.~Golbraikh and A.~Tropsha, ``Predictive qsar modeling based on diversity
  sampling of experimental datasets for the training and test set selection,''
  \emph{Molecular diversity}, vol.~5, no.~4, pp. 231--243, 2000.

\bibitem{gokcesu2020generalized}
K.~Gokcesu and H.~Gokcesu, ``A generalized online algorithm for translation and
  scale invariant prediction with expert advice,'' \emph{arXiv preprint
  arXiv:2009.04372}, 2020.

\bibitem{fan2005working}
R.-E. Fan, P.-H. Chen, C.-J. Lin, and T.~Joachims, ``Working set selection
  using second order information for training support vector machines.''
  \emph{Journal of machine learning research}, vol.~6, no.~12, 2005.

\bibitem{huberbook}
T.~Hastie, R.~Tibshirani, and J.~Friedman, \emph{The Elements of Statistical
  Learning}, ser. Springer Series in Statistics.\hskip 1em plus 0.5em minus
  0.4em\relax New York, NY, USA: Springer New York Inc., 2001.

\bibitem{cesabook}
N.~Cesa-Bianchi and G.~Lugosi, \emph{Prediction, learning, and games}.\hskip
  1em plus 0.5em minus 0.4em\relax Cambridge university press, 2006.

\bibitem{poor_book}
H.~V. Poor, \emph{An Introduction to Signal Detection and Estimation}.\hskip
  1em plus 0.5em minus 0.4em\relax NJ: Springer, 1994.

\bibitem{kellerer2004knapsack}
H.~Kellerer, U.~Pferschy, and D.~Pisinger, \emph{Knapsack problems.}\hskip 1em
  plus 0.5em minus 0.4em\relax Springer, 2004.

\bibitem{gokcesu2022nonconvex}
K.~Gokcesu and H.~Gokcesu, ``Nonconvex extension of generalized huber loss for
  robust learning and pseudo-mode statistics,'' \emph{arXiv preprint
  arXiv:2202.11141}, 2022.

\bibitem{mathews1896partition}
G.~B. Mathews, ``On the partition of numbers,'' \emph{Proceedings of the London
  Mathematical Society}, vol.~1, no.~1, pp. 486--490, 1896.

\bibitem{dantzig2007number}
T.~Dantzig, \emph{Number: The language of science}.\hskip 1em plus 0.5em minus
  0.4em\relax Penguin, 2007.

\bibitem{coffman1984approximation}
E.~G. Coffman, M.~R. Garey, and D.~S. Johnson, ``Approximation algorithms for
  bin-packing—an updated survey,'' in \emph{Algorithm design for computer
  system design}.\hskip 1em plus 0.5em minus 0.4em\relax Springer, 1984, pp.
  49--106.

\bibitem{brams1996fair}
S.~J. Brams, S.~J. Brams, and A.~D. Taylor, \emph{Fair Division: From
  cake-cutting to dispute resolution}.\hskip 1em plus 0.5em minus 0.4em\relax
  Cambridge University Press, 1996.

\bibitem{biswas2018fair}
A.~Biswas and S.~Barman, ``Fair division under cardinality constraints.'' in
  \emph{IJCAI}, 2018, pp. 91--97.

\bibitem{gokcesu2021optimal}
K.~Gokcesu and H.~Gokcesu, ``Optimal and efficient algorithms for general
  mixable losses against switching oracles,'' \emph{arXiv preprint
  arXiv:2108.06411}, 2021.

\bibitem{walsh2009really}
T.~Walsh, ``Where are the really hard manipulation problems? the phase
  transition in manipulating the veto rule,'' in \emph{Twenty-First
  International Joint Conference on Artificial Intelligence}, 2009.

\bibitem{merkle1978hiding}
R.~Merkle and M.~Hellman, ``Hiding information and signatures in trapdoor
  knapsacks,'' \emph{IEEE transactions on Information Theory}, vol.~24, no.~5,
  pp. 525--530, 1978.

\bibitem{shamir1982polynomial}
A.~Shamir, ``A polynomial time algorithm for breaking the basic merkle-hellman
  cryptosystem,'' in \emph{23rd Annual Symposium on Foundations of Computer
  Science (sfcs 1982)}.\hskip 1em plus 0.5em minus 0.4em\relax IEEE, 1982, pp.
  145--152.

\bibitem{rivest1983cryptographic}
R.~L. Rivest, A.~Shamir, and L.~M. Adleman, ``Cryptographic communications
  system and method,'' Sep.~20 1983, uS Patent 4,405,829.

\bibitem{gokcesu2021optimally}
K.~Gokcesu and H.~Gokcesu, ``Optimally efficient sequential calibration of
  binary classifiers to minimize classification error,'' \emph{arXiv preprint
  arXiv:2108.08780}, 2021.

\bibitem{sarkar1987partitioning}
V.~Sarkar, ``Partitioning and scheduling parallel programs for execution on
  multiprocessors,'' Ph.D. dissertation, Stanford University, 1987.

\bibitem{dell2008heuristic}
M.~Dell'Amico, M.~Iori, S.~Martello, and M.~Monaci, ``Heuristic and exact
  algorithms for the identical parallel machine scheduling problem,''
  \emph{INFORMS Journal on Computing}, vol.~20, no.~3, pp. 333--344, 2008.

\bibitem{graham1979optimization}
R.~L. Graham, E.~L. Lawler, J.~K. Lenstra, and A.~R. Kan, ``Optimization and
  approximation in deterministic sequencing and scheduling: a survey,'' in
  \emph{Annals of discrete mathematics}.\hskip 1em plus 0.5em minus 0.4em\relax
  Elsevier, 1979, vol.~5, pp. 287--326.

\bibitem{umang2013exact}
N.~Umang, M.~Bierlaire, and I.~Vacca, ``Exact and heuristic methods to solve
  the berth allocation problem in bulk ports,'' \emph{Transportation Research
  Part E: Logistics and Transportation Review}, vol.~54, pp. 14--31, 2013.

\bibitem{lalla2016set}
E.~Lalla-Ruiz, C.~Exp{\'o}sito-Izquierdo, B.~Meli{\'a}n-Batista, and J.~M.
  Moreno-Vega, ``A set-partitioning-based model for the berth allocation
  problem under time-dependent limitations,'' \emph{European Journal of
  Operational Research}, vol. 250, no.~3, pp. 1001--1012, 2016.

\bibitem{gokcesu2022low}
K.~Gokcesu and H.~Gokcesu, ``Low regret binary sampling method for efficient
  global optimization of univariate functions,'' \emph{arXiv preprint
  arXiv:2201.07164}, 2022.

\bibitem{horowitz1974computing}
E.~Horowitz and S.~Sahni, ``Computing partitions with applications to the
  knapsack problem,'' \emph{Journal of the ACM (JACM)}, vol.~21, no.~2, pp.
  277--292, 1974.

\bibitem{schroeppel1981t}
R.~Schroeppel and A.~Shamir, ``A {$T=O\left(2^{n/2}\right)$},
  {$S=O\left(2^{n/4}\right)$} algorithm for certain np-complete problems,''
  \emph{SIAM journal on Computing}, vol.~10, no.~3, pp. 456--464, 1981.

\bibitem{martello1990knapsack}
S.~Martello, ``Knapsack problems: algorithms and computer implementations,''
  \emph{Wiley-Interscience series in discrete mathematics and optimiza tion},
  1990.

\bibitem{korf2013optimally}
R.~E. Korf and E.~L. Schreiber, ``Optimally scheduling small numbers of
  identical parallel machines,'' in \emph{Twenty-Third International Conference
  on Automated Planning and Scheduling}, 2013.

\bibitem{karmarkar1982differencing}
N.~Karmarkar and R.~M. Karp, \emph{The differencing method of set
  partitioning}.\hskip 1em plus 0.5em minus 0.4em\relax Computer Science
  Division (EECS), University of California Berkeley, 1982.

\bibitem{korf2011hybrid}
R.~E. Korf, ``A hybrid recursive multi-way number partitioning algorithm,'' in
  \emph{Twenty-Second International Joint Conference on Artificial
  Intelligence}, 2011.

\bibitem{gokcesu2021efficient}
K.~Gokcesu and H.~Gokcesu, ``Efficient locally optimal number set partitioning
  for scheduling, allocation and fair selection,'' \emph{arXiv preprint
  arXiv:2109.04809}, 2021.

\bibitem{gokcesu2021quadratic}
------, ``A quadratic time locally optimal algorithm for np-hard equal
  cardinality partition optimization,'' \emph{arXiv preprint arXiv:2109.07882},
  2021.

\bibitem{gokcesu2022linearithmic}
------, ``A linearithmic time locally optimal algorithm for the multiway number
  partition optimization,'' \emph{arXiv preprint arXiv:2203.05618}, 2022.

\end{thebibliography}

\end{document}